\documentclass[letterpaper, 10 pt, conference]{ieeeconf}

\usepackage{graphicx}
\usepackage{balance}
\usepackage{comment}
\usepackage{cite}
\usepackage{amssymb}
\usepackage[tight,footnotesize]{subfigure}
\usepackage[active]{srcltx}
\usepackage{amsmath}

\newcommand{\norm}[1]{\left\lVert#1\right\rVert}
\graphicspath{{./figs/}}
\renewcommand{\baselinestretch}{0.95}
\usepackage{eurosym}
\usepackage{algorithm}
\usepackage{algorithmic}
\usepackage{multicol}
\usepackage{dsfont}
\usepackage{amsfonts}
\newcommand{\normvec}[1]{\left\lVert#1\right\rVert}
\usepackage{cases}
\IEEEoverridecommandlockouts
\begin{document}

\title{Decentralized Search and Track with Multiple Autonomous Agents}

\author{Savvas~Papaioannou,~Panayiotis~Kolios,~Theocharis~Theocharides,\\~Christos~G.~Panayiotou~ and ~Marios~M.~Polycarpou
\thanks{The authors are with the KIOS Research and Innovation Centre of Excellence (KIOS CoE) and the Department of Electrical and Computer Engineering, University of Cyprus, Nicosia, 1678, Cyprus. {\tt\small \{papaioannou.savvas, pkolios, ttheocharides, christosp, mpolycar\}@ucy.ac.cy}}
}

\maketitle

\begin{abstract}
In this paper we study the problem of cooperative searching and tracking (SAT) of multiple moving targets with a group of autonomous mobile agents that exhibit limited sensing capabilities. We assume that the actual number of targets is not known a priori and that target births/deaths can occur anywhere inside the surveillance region. For this reason efficient search strategies are required to detect and track as many targets as possible. To address the aforementioned challenges we augment the classical Probability Hypothesis Density (PHD) filter with the ability to propagate in time the search density in addition to the target density. Based on this, we develop decentralized cooperative look-ahead strategies for efficient searching and tracking of an unknown number of targets inside a bounded surveillance area. The performance of the proposed approach is demonstrated through simulation experiments.
\end{abstract}

%
\section{Introduction}
\label{sec:Introduction}
One of the biggest challenges today’s society faces is its resilience to severe disasters. Unfortunately, first responders currently rely on a number of conventional methods to gather information that are time consuming and the descriptive character of the collected information often lacks accuracy, eloquence and the necessary level of detail. In this work, we envision that a team of autonomous mobile agents (e.g. drones) could become an important technological tool to aid the work of the rescuers. Under this setting, the mission of one or more drone agents is to assist first responders by conducting the following important tasks: a) search the area for situational assessment, and b) detect and track victims as accurately as possible. More specifically, in a cooperative search and track (SAT) mission, multiple agents are tasked to cooperatively search a certain area of interest in order to discover survivors while at the same time keeping track of those survivors already detected. 
This work builds upon the theory of random finite sets (RFS) and proposes a multi-agent 
framework for SAT missions that takes into account the unknown and time varying number of survivors, the noisy sensor measurements and the limited sensing range of the agents. In addition, efficient cooperative search and track strategies are devised which allow the agents to generate joint search-plans and detect and resolve tracking overlaps. The contributions of this paper are as follows:
\begin{itemize}
	\item Devises efficient cooperative searching and tracking strategies for a decentralized multi-agent framework.
    \item Provides a new perspective on the problem of multi-agent cooperative searching and tracking (SAT) through a unified probabilistic approach based on random finite sets (RFS).
    \item Proposes a method to recursively compute and propagate in time the SAT-density by extending the classical probability hypothesis density (PHD) filter to account for the search density in addition to the target density.  	
\end{itemize}
 
\section{Related Work}\label{sec:Related_Work}
Previous works in \cite{Bourgault2003} and \cite{Liu2017} investigate the SAT problem but only for the single-agent single-target case. The work in \cite{Furukawa2006} proposes a recursive Bayesian multi-agent SAT solution, however the agents are required to be in communication range at all times. The work in \cite{Frew2008} proposes a task assignment algorithm that integrates area search and target tracking, however requires that the number of agents is larger than the number of targets and that a single agent can only track one target at a time. The problem of multi-agent SAT is also investigated in \cite{Pitre2012} but lacks online path generation. Finally, the work in \cite{Peterson2017} proposes a cooperative search and track framework, and a clustering approach for grouping neighboring agents (that have intersecting decision spaces) in order to minimize complexity. This work however, assumes clutter free (i.e. no false-alarm measurements are received) environment, perfect target detection and that targets can be uniquely identified. 
Relevant works also include, the work in \cite{Moratuwage2013} which presents an interesting use of random finite sets on collaborative multi-vehicle SLAM and the works in \cite{Dames2017,Dames2019} which implement efficient multi-agent RFS-based simultaneous coverage and tracking algorithms for tracking multiple targets.

Complimentary to the related work, in this paper we propose a decentralized architecture where multiple agents cooperatively search a region of interest, detect targets in the area and perform tracking of multiple detected targets.We assume that a particular 2D region of interest needs to be continuously searched for potential targets with the aid of a group of mobile agents. The number of targets is not known a priori and changes over time. The agents are equipped with sensors and receive noisy measurements from the targets in the presence of clutter (i.e. false-alarm measurements). The agents have a limited sensing range for detecting targets and limited communication range for exchanging information with other nearby agents. Importantly, the aforementioned problem has been identified as the hardest version of the SAT problem that has not been addressed adequately in the literature as indicated in \cite{Chung2018,Khan2018}.

That said, the objective of each agent at an arbitrary time-step is to: a) accurately estimate the number of targets and their states inside its sensing range from noisy measurements in the presence of clutter, and b) generate search-plans for efficiently searching the whole surveillance area. To achieve a) and b), each agent maintains a modified PHD filter, termed in this paper as SAT-PHD filter, which in addition to the target density, it recursively computes the search density. Finally, the agents opportunistically cooperate by exchanging information in order to tackle the above objectives more efficiently e.g. when two or more agents are in communication range they cooperate to generate joint search-plans and to resolve tracking overlaps (i.e. a situation where 2 or more agents track the same targets). To summarize, the agents in communication range exchange their search densities, their multi-target states and their mode of operation i.e. search or track. We should also note that all agents are in search mode optimizing their local or joint search objective (see subsection \ref{ssec:cooperative_search}) until targets are found in the surveillance area in which case the respective agents switch to track mode (see subsection \ref{ssec:multi_agent_tracking}).


\section{Background on Random Finite Sets} \label{sec:Background}

The goal of Bayesian filtering \cite{Chen2003} is to recursively estimate the conditional posterior distribution i.e. $p(x_k|z_{1:k})$, of the target state $x_k$ at time $k$ based on the history of measurements $z_1,z_2,...,z_k$ up to time $k$. In the single target tracking scenario the target state $x_k$ and measurement $z_k$ can be represented as random variables or random vectors with fixed size i.e. the state of a target (e.g. position) can change over time however the dimension of the state vector remains constant.
On the other hand, in a multi-target system the number of targets changes over time as targets enter and exit the surveillance area which results in a multi-target state (i.e. a collection of individual target states) that changes size over time i.e. the dimension of this multi-target state varies over time as opposed to the dimension of the single target state which remains constant. Using the theory of random finite sets (RFSs) \cite{Mahler2014book} the collection of target states can be represented as finite subsets $X_k = \{x^1_k, x^2_k,...,x^{n_k}_k\} ~ \in \mathcal{F}(\mathcal{X})$ where $\mathcal{X}$ denotes the state-space of the single target state and $\mathcal{F(X)}$ denotes the space of all finite subsets of $\mathcal{X}$. Finally $n_k$ is the true but unknown number of targets that needs to be estimated. The set $X_k$ is called random finite set and can be seen as a generalization of a random vector. 

The multi-object conditional distribution $f_k(X_k|Z_{1:k})$ of the RFS $X_k$ based on measurements $Z_{1:k}$ up to time $k$ can be estimated using Bayesian multi-object stochastic filtering. However, the optimal multi-object Bayes filter is in general intractable and has no analytical solution. A practical alternative is the Probability Hypothesis Density (PHD) filter \cite{Mahler2003} which only propagates the first-order statistical moment instead of the full multi-object posterior distribution. More specifically, the PHD at time $k$ is the conditional density $D_k(x|Z_{1:k})$ which when integrated over any region $R \subseteq \mathcal{X}$ gives the expected number of targets $\hat{n}_k$ contained in $R$, i.e. $\hat{n}_k(R) = \int_{R} D_k(x|Z_{1:k}) dx$, where the notion of integration is given by the set-integral \cite{Mahler2014book}. Finally, the multi-target state $\hat{X}_k$ can be estimated as the $\hat{n}_k$ highest local maxima of the PHD. 

%

\section{System Model} \label{sec:system_model}
%
\subsection{Single Target Dynamics and Measurement Model}\label{ssec:single_target_dynamics}
Let the state of a single target have the following form:
\begin{equation}
   (x,\ell) \in \mathcal{X} \times \{0,1\}
\end{equation}
where $x \in \mathcal{X}$ is the kinematic state of the target, $\mathcal{X} \subseteq R^{n_x}$ denotes the kinematic state space of the target, $n_x$ is the dimension of the state vector $x$ and $\ell \in \{0,1\}$ is the target label taken from the discrete label space $\{0,1\}$. We denote a true target with label $\ell=1$ and a virtual target with label $\ell=0$. True targets represent physical targets inside the surveillance region whose kinematic state $x$ needs to be estimated from a sequence of noisy measurements whereas virtual targets represent static locations in the environment which will be used to model the state of searching i.e. whether or not these locations have been searched. Throughout this paper, the kinematic state spaces of true and virtual targets will be denoted as $\mathcal{X}^1$ and $\mathcal{X}^0$, respectively. The single target kinematic state vector $x_k, k \in \mathbb{N}$  evolves in time according to the following equation:
\begin{subnumcases}{x_k=}
   \zeta(x_{k-1}) + w_k & \text{, if} $~x_{k-1} \in \mathcal{X}^1$ \label{eq:single_dynamics_a} \\
   x_{k-1} & \text{, if} $~x_{k-1} \in \mathcal{X}^0$ \label{eq:single_dynamics_b}
\end{subnumcases}

\noindent where the function $\zeta : \mathbb{R}^{n_x} \rightarrow \mathbb{R}^{n_x}$ models the dynamical behavior of the target. Eqn. (\ref{eq:single_dynamics_a}) describes the evolution of the state vector as a first order Markov process with transitional density $\pi_{k|k-1}(x_k|x_{k-1}) = p_w(x_k - \zeta(x_{k-1}) )$. The process noise $w_{k} \in \mathbb{R}^{n_x}$ is independent and identically distributed (IID) according to the probability density function $p_w(.)$. In this paper we assume that the kinematic state vector $x_k \in \mathcal{X} \subseteq \mathbb{R}^4$ is composed of position and velocity components in Cartesian coordinates i.e. $x_k = [\text{x},\dot{\text{x}},\text{y},\dot{\text{y}}]^\top$. Since a virtual target is static, its kinematic state vector is of the form $x_k = [\text{x},0,\text{y},0]^\top$. When an agent detects a true target i.e. $x_k \in \mathcal{X}^1$ at time $k$, it receives a measurement vector $z_k \in \mathcal{Z}$ (range and bearing observations) which is related to the target kinematic state as follows:
\begin{equation}
    z_k = h(x_k,s_k) + \text{v}_k
\end{equation}
where $\mathcal{Z} \subseteq \mathbb{R}^{n_z}$ denotes the measurement space, $s_k$ is the state of the agent at time $k$ (described in the next sub-section) and the function $h(.,.)$ projects the state vector to the measurement space. The random process $\text{v}_{k} \in \mathbb{R}^{n_z}$ is IID, independent of $w_k$ and distributed according to $p_\text{v}(.)$. The probability density of measurement $z_k$ for a target with kinematic state $x_k$ when the agent is at state $s_k$ is given by the measurement likelihood function $g_k(z_k|x_k,s_k) = p_\text{v}(z_k - h(x_k,s_k))$. On the other hand, virtual targets are observed directly without noise i.e. the measurement of a virtual target is its actual state. 

\subsection{Agent Dynamics}
\label{ssec:AgentDynamics}
Let $S = \{1,2,...,|S|\}$ be the set of all mobile agents that we have in our disposal operating in a discrete-time setting. At time $k$, the 2D surveillance region $\mathcal{A} \subseteq \mathbb{R}^2$ is monitored by $|S|$ mobile agents with states $s^1_k,s^2_k,...,s^{|S|}_k$, each taking values in $\mathcal{A}$. Each agent $j$ is subject to the following dynamics:
\begin{equation} \label{eq:controlVectors}
s^j_{k} = s^j_{k-1} + \begin{bmatrix}
						l_1\Delta_R \text{cos}(l_2 \Delta_\theta)\\
						l_1\Delta_R \text{sin}(l_2 \Delta_\theta)
					\end{bmatrix},  
					\begin{array}{l} 
						l_2 = 0,...,N_\theta\\ 
						l_1 = 0,...,N_R
				    \end{array} 
\end{equation}
where  $s^j_{k-1} = [s^j_x,s^j_y]^\top_{k-1}$ denotes the position (i.e. xy-coordinates) of the $j_{\text{th}}$ agent at time $k-1$, $\Delta_R$ is the radial step size, $\Delta_\theta=2\pi/N_\theta$ and the parameters $(N_\theta,N_R)$ control the number of possible control actions. We denote the set of all admissible control actions of agent $j$ at time $k$ as $\mathbb{U}^j_{k}=\{s^{j,1}_{k},s^{j,2}_{k},...,s^{j,|\mathbb{U}_{k}|}_{k}  \}$ as computed by Eqn. (\ref{eq:controlVectors}). 

\subsection{Single Agent Sensing Model} \label{ssec:measM}
The ability of an agent to sense its 2D environment is modeled by the function $p_D(x_k,s_k)$ that measures the probability that a target with kinematic state $x_k$ at time $k$ is detected by an agent with state $s_k$. More specifically, when $x_k \in \mathcal{X}^1$ the sensing capability of the agent is given by:
\begin{equation}\label{eq:sensing_model}
 p_D(x_k \in \mathcal{X}^1 ,s_k) = 
  \begin{cases} 
   p_D^\text{max} & \text{, if } x_k \in \mathcal{S}_a(s_k) \\
   0 & \text{, if } x_k \notin \mathcal{S}_a(s_k)
  \end{cases}
\end{equation}
where $\mathcal{S}_a(s_k)$ denotes the agent's sensing area which in this work includes all $xy$ points that satisfy the equation $\max \{|x-s_x|,|y-s_y|\}=\frac{a}{2}$, i.e. a square region with total area $a^2$ units, centered at $s_k = [s_x, s_y]^\top$ and $p_D^\text{max}$ denotes the probability of the sensor to detect true targets inside its sensing range. On the other hand, the agent detects virtual target inside its sensing range with probability 1 i.e. $p_D(x_k \in \mathcal{X}^0 ,s_k) = 1$ when $x_k \in \mathcal{S}_a(s_k)$ and $p_D(x_k \in \mathcal{X}^0 ,s_k) = 0$ when $x_k \notin \mathcal{S}_a(s_k)$. In addition, any two agents with states $s^i_k$ and $s^j_k$ are able to communicate with each other when $\normvec{s^i_k - s^j_k}_2 \le C_R$ where $C_R$ is the communication range.

\subsection{Multi-object dynamics and measurement models}
Multiple independent targets can exist and evolve inside the surveillance region. True targets (i.e. with label $\ell=1$) can spawn from anywhere in the state space $\mathcal{X}^1$ and target births and deaths occur at random times. This means that at each time $k$, there exist $n^{\ell=1}_k$ true targets with kinematic states $x^1_k, x^2_k,...,x^{n^{\ell=1}_k}_k$, each taking values in the state space $\mathcal{X}^1$ where both the number of true targets $n^{\ell=1}_k$ and their individual states $x_k^i, \forall i \in n^{\ell=1}_k$ are random and time-varying. The multi-target (or multi-object) state of the true targets is thus represented as the RFS  $X^{\ell=1}_k \in \mathcal{F(X^\text{1})}$ which evolves in time according to: $X^{\ell=1}_k =  \underset{x_{k-1} \in X^{\ell=1}_{k-1}}{\bigcup} \Psi(x_{k-1})  \cup B_k$

\noindent where $X^{\ell=1}_{k-1}$ is the multi-target state of the true targets of previous time-step, $\Psi(x_{k-1})$ is a Bernoulli RFS \cite{Ristic2013} which models the evolution of the set from the previous state, with parameters $(p_{S}(x_{k-1}),\pi_{k|k-1}(x_k|x_{k-1}))$. Thus a target with kinematic state $x_{k-1}$ continues to exists at time $k$ with surviving probability $p_{S}(x_{k-1})$ and moves to a new state $x_k$ with transition probability $\pi_{k|k-1}(x_k|x_{k-1})$. Otherwise, the target dies with probability $1-p_{S}(x_{k-1})$. The term $B_k$ is the RFS of spontaneous births \cite{Mahler2003}.

The virtual targets i.e. $(\ell=0)$ on the other hand do not exhibit any birth and death events, and their number $n^{\ell=0}_k=n^{\ell=0}$ is constant and known (i.e. sampled uniformly from the surveillance area at $k=0$). Thus the multi-target state at time $k$ is given by $X_k = X^{\ell=1}_k \cup X^{\ell=0}_k$ where $X^{\ell=0}_k$ is the set of all virtual targets in the surveillance region with $|X^{\ell=0}_k| = n^{\ell=0} ~ \forall ~ k$. In the rest of the paper we abbreviate $X^{\ell=1}_k =X^1_k$ and $X^{\ell=0}_k =X^0_k$. At time $k$, an agent receives a finite set of measurements (i.e. measurements generated from the detected true targets and from clutter) denoted as $Z_k$. This RFS has the form: $Z_k =  \underset{x_{k} \in X^1_{k}}{\bigcup} \Theta(x_{k})  \cup \text{K}_k$ where $\Theta(x_{k})$ is a Bernoulli RFS which models the target generated measurements with parameters $(p_{D}(x_k,s_k),g_k(z_k|x_k,s_k))$. Thus a true target with kinematic state $x_k$ at time $k$ is detected by the agent with state $s_k$ with probability $p_{D}(x_k,s_k)$ and receives a measurement $z_k$ with likelihood $g_k(z_k|x_k,s_k)$ or is missed with probability $1-p_{D}(x_k,s_k)$ and generates no measurements. Additionally, an agent can receive false alarms measurements i.e. the term $\text{K}_k$ is a Poisson RFS which models the set of false alarms or clutter received by an agent at time $k$ with PHD $\kappa_k(z_k) = \lambda f_{c}(z_k)$, where in this paper $f_c(.)$ denotes the uniform distribution over $\mathcal{Z}$ and $\lambda$ is the average number of clutter generated measurements per time-step.

\section{Proposed Approach} \label{sec:proposed_approach}
In this section we describe how we have extended the classical PHD filter to propagate in time the SAT-density and next we discuss how using the SAT-density the agents cooperate to produce joint search-plans and resolve tracking overlaps. 

\subsection{Search-and-Track Density}
During a search and track mission, a single agent is required to be able to perform the following tasks: a) simultaneously estimating the time-varying number of targets and their states from a sequence of noisy measurements and b) efficiently searching the surveillance region in order to maximize the probability of finding targets.

The first task can be accomplished by recursively computing and propagating in time, the PHD of the full multi-target posterior distribution using the PHD filter \cite{Mahler2003}. In order to accomplish the second task, the agent needs to: a) keep track of the visited (i.e. searched) and unvisited regions of the surveillance area, b) intelligently estimate when and how often certain search regions need to be revisited (i.e. searched again), and c) generate efficient search plans for searching the area. To do that we assume that the agent stores a discrete representation of the environment in its memory in the form of a graph $G=\{\mathcal{V},\mathcal{E}\}$ termed as \textit{search map}, where each node $v \in \mathcal{V}$ corresponds to a region $r_v \subset \mathcal{A}$ in the surveillance area where $\bigcup_v r_v = \mathcal{A}$. The agent recursively computes the \textit{search value} $p^\text{search}(r_v) \in [0,1], v\in \mathcal{V}$ of each region and uses this information to decide how often to visit a particular region and how to generate search-plans for efficiently searching the surveillance area. 

With this in mind, we have extended the classic PHD filter in order to recursively compute the search density in addition to the target density. At each time-step we use the target density to estimate the number of targets inside the agents' sensing range and the search density to compute the search values of every region in the surveillance area. More specifically, the predicted SAT-PHD at $x \in \mathcal{X}$ can be computed as:
\begin{align}\label{eq:satphd_predict}
& D_{k|k-1}(x|Z_{1:k-1}) = b_{k}(x\in \mathcal{X}^1) ~+ \notag \\ 
&\int_{\mathcal{X}^1} p_S(x^\prime) \pi_{k|k-1}(x\in \mathcal{X}^1|x^\prime) D_{k-1}(x^\prime|Z_{1:k-1}) d x^\prime +  \\
& \Big[ \big(1 - p_D(x\in \mathcal{X}^0,s_{k-1})\big)  J_k(x\in \mathcal{X}^0) + p_D(x\in \mathcal{X}^0,s_{k-1}) \Big]  \notag \\
& \cdot D_{k-1}(x\in \mathcal{X}^0|Z_{1:k-1})\notag
\end{align}
\noindent where $b_{k}(x)$ is the PHD of target births, $p_S(x)$ is the probability that a target with state $x$ will survive in the next time step and $\pi_{k|k-1}(x|x^\prime)$ is the single-target transition density, $p_D(x,s_k)$ is the sensing model defined in Eqn. (\ref{eq:sensing_model}) and $J_k(x) \in [0..1]$ is a function that determines the decay value of the virtual target with state $x$. The first two lines of Eqn. (\ref{eq:satphd_predict}) are due to the classic PHD filter which are used to predict the target density at $x$, whereas the 3rd line is used to predict the search density operating only on virtual targets. In essence, the states of all virtual targets outside the agent's sensing range are adjusted accordingly to reflect the fact that they are not being observed. This property is used to generate search plans which will guide the agent to visit areas that have not been recently visited. The updated SAT-PHD density is given by:
\begin{align} \label{eq:satphd_correct}
& D_{k}(x|Z_{1:k}) = \!\!\Big[1 - p_D(x\in \mathcal{X}^1\!\!,s_k) \Big]D_{k|k-1}(x\in \mathcal{X}^1|Z_{1:k-1}) ~   \notag \\
& ~~~ + \Bigg[ \underset{z \in Z_k}{\sum} \frac{p_D(x\in \mathcal{X}^1,s_k) \cdot g_k(z|x\in \mathcal{X}^1,s_k)}{\kappa(z)+\tau(z)} \Bigg]  ~ \cdot \notag \\
& ~~~~ D_{k|k-1}(x\in \mathcal{X}^1|Z_{1:k-1}) + \frac{p_D(x\in \mathcal{X}^0,s_k)}{|\mathcal{A}|} \\
& ~~~~+  \Big[1-p_D(x\in \mathcal{X}^0,s_k)\Big]  D_{k|k-1}(x\in \mathcal{X}^0|Z_{1:k-1})  \notag
\end{align}

\noindent where $|\mathcal{A}|$ is the total area of the surveillance region and $\tau(z) = \int_{\mathcal{X}^1} p_D(x^\prime,s_k) g_k(z|x^\prime,s_k) D_{k|k-1}(x^\prime|Z_{1:k-1}) d x^\prime $. In the above equation the last term was added to the classical PHD filter update step in order to adjust the search density inside the agents sensing range to account for the agent's updated position $s_k$. Finally, the search value $p^\text{search}_k(r_v)$ of a particular region $r_v \subset \mathcal{A},\>v\in\mathcal{V}$ can be computed by integrating the SAT-PHD in $r_v$ as follows:
\begin{equation} \label{eq:search_value}
p^\text{search}_k(r_v)=\frac{\int_{r_v}D_k(x \in \mathcal{X}^0 |Z_{1:k}) dx}{|r_v||\mathcal{A}|^{-1}}
\end{equation}
where $|r_v|$ is the area of region $r_v$. Finally, the number of true targets $\hat{n}_k$ inside the area $R \subseteq \mathcal{A}$ can be computed by integrating the SAT-PHD in $R$ as $\hat{n}_k(R) = \int_{R} D_k(x \in \mathcal{X}^1|Z_{1:k}) dx$ (rounded to the nearest integer) and the multi-target state $X^1_k$ can be estimated by finding the $\hat{n}_k(R)$ highest peaks of the PHD as in the original PHD filter.

\subsection{Multi-agent Searching} \label{ssec:cooperative_search}
The search objective is to find the optimal control actions that will move the agent along areas that have not been explored for some time and could potentially reveal new targets. To address this challenge, we first discuss how searching takes into account the search map derived from the SAT-PHD filter and how low level controls employ the computed paths to steer the agent across the field.

\textbf{a) Search planning:} Given the search map $G=(\mathcal{V},\mathcal{E})$ where the set of edges in $\mathcal{E}$ connect adjacent nodes, the cost $c_{ij}$ on edge $i\mapsto j$ is defined as the Euclidean distance between the particular regions in the field. For each node on this graph, a search value $p^{\text{search}}(r_v),\>v\in\mathcal{V}$ is computed using Eqn. (\ref{eq:search_value}). This value varies between 0 and 1, where 0 indicates that the particular node (and hence region in the field) has not been searched and 1 indicates that the region has just been visited. The SAT-PHD recursion in Eqn.(\ref{eq:satphd_predict}) - (\ref{eq:satphd_correct}) indicates how the search value decays over time in order to steer agents to revisit particular regions in the field.

Using $p^\text{search}$, we then define the set of unvisited nodes $\bar{\mathcal{V}}$ as the set of nodes for which the search value goes below a certain threshold, i.e., $p^\text{search}(r_v)\leq\beta, v\in\bar{\mathcal{V}}$ and thus indicating that those nodes need to be revisited. Given $\bar{\mathcal{V}}$ and the initial location $s_k$ of agent $k$, we would like to compute closed walks starting at $s_k$ to visit nodes in $\bar{\mathcal{V}}$ with the least cost $c_{ij}$ in order to search the field for targets. Note that computing optimal closed walks with the least cost can be achieved by employing variations of the vehicle routing problem. However due to the high computationally complexity of the optimal algorithms, this approach can not be employed in practice. Hence, an alternative heuristic approach is followed hereafter to devise closed walks efficiently in time. Similar to the path-cheapest-arc heuristic \cite{ORTools}, each walk is set to start at $s_k$, and a path is constructed greedily by inserting each new edge with the least cost emanating from the head node of the last edge added. The process repeats until all nodes have been visited or when no more edges can be added.

\textbf{b) Control:} Given the computed closed-walk sequence, the objective is then to take a control action $u_k \in \mathbb{U}_k$ that will move the agent across the designated path. To achieve this, a list of nodes to-be-visited is maintained, and each node is marked as visited whenever the agent moves to a position where the particular node is within its sensing range. The control objective can simply be expressed as follows, $u_{k}^{\star} = \arg\min~ \xi_\text{search}(u_{k},v)$, where $v\in\bar{\mathcal{V}}$ indicates the location of the next unvisited node in the list and $\xi_\text{search}$ returns the Euclidean distance between the current position of the agent and the next unvisited node. By iteratively visiting the close-walk sequence, the envisioned look-ahead search control is achieved by each agent.

\textbf{c) Cooperation:} Whenever two or more agents are in communication range they exchange their search densities. Agents then merge their copies using a simple max operation of local and received values and compute a fused search map which now contains the search-path histories of the involved agents. The agents can then compute a joint search-plan as follows: Let $\bar{\mathcal{S}}$ be the subset of agents in communication range and assume that each agent knows the number $|\bar{\mathcal{S}}|$ and position $s_k^j,\>j\in \bar{\mathcal{S}}$ of cooperating agents in its vicinity. Then iteratively each agent computes $|\bar{\mathcal{S}}|$ closed walks incrementally by adding one node at a time in each agent's path from the list of all unvisited nodes in $\mathcal{V}$ until there are no more unvisited nodes. A new node is added in an agent's path only if the head node of all possible edges to traverse (starting from the edge with the least cost) is not flagged as visited and the tail node of that edge is the last node added in the particular walk.

\subsection{Multi-agent Tracking} \label{ssec:multi_agent_tracking}
In this subsection we discuss: a) how multiple agents are cooperating to detect and resolve tracking overlaps and b) how the agents select control actions in order to accurately track multiple targets.

\textbf{a) Tracking overlap detection:} This problem arises when a target is being tracked by more than one agent. This is something unwanted since valuable system resources are wasted for performing the same task. Consider, the scenario where 3 targets, which are being tracked by two different agents, approach each other. 
Eventually, the targets will be detected and tracked by both agents at the same time. We denote this situation as a tracking overlap event, which we wish to detect and resolve. To do so, and instead of solving the combinatorial problem that arises (which requires the enumeration of joint control actions among agents and future multi-target states over a finite horizon), in this work we propose a computationally cheaper way to tackle the tracking overlap problem.

 More specifically, in order to reduce the computational and communication overhead, we allow any two agents to merge and track the same targets but only for a short period of time. More specifically, we consider that each agent can track multiple targets independent of other agents. When, the trajectories of two or more tracking agents converge the agents exchange information to determine whether or not the exact same targets are being tracked. Once, two agents have determined that they track exactly the same targets, one of them generates a search plan and exits the tracking. The above procedure begins when two or more tracking agents have overlapping sensing ranges. 

Let the predicted multi-target states (regarding the true targets) of any two agents with states $s^i_{k-1}$ and $s^j_{k-1}$ be $\hat{X}^{1,i}_{k|k-1}$ and $\hat{X}^{1,j}_{k|k-1}$, respectively.
The predicted multi-target state $\hat{X}^{1}_{k|k-1}$ is computed from the predicted SAT-PHD i.e. Eqn. (\ref{eq:satphd_predict}) by selecting the $\hat{n}_{k|k-1}$ highest peaks where $\hat{n}_{k|k-1} = \int D_{k|k-1}(x \in \mathcal{X}^1|Z_{1:k}) dx $. Also, let  $|\hat{X}^{1,i}_{k|k-1}| = m$ and $|\hat{X}^{1,j}_{k|k-1}| = n$ denote their cardinalities, i.e. the number of predicted targets in the set, with $n \ge m$ and $n, m \ne 0$. When $\mathcal{S}_a(s^i_{k-1}) \cap \mathcal{S}_a(s^j_{k-1}) \neq \emptyset$, the agents exchange their predicted multi-target states to compute the \textit{incremental tracking overlap score} as:
\begin{align} \label{eq:ospa}
&\Delta L^c_{k}(\hat{X}^{1,i}_{k|k-1},\hat{X}^{1,j}_{k|k-1}) = \notag \\
&\>\>\> \Bigg[~\frac{1}{n} \Bigg(\underset{\pi \in \Pi_n}{\text{min}} ~\underset{l=1}{\sum^m} ~ d_c(x^i_l,x^j_{\pi(l)})^2 + (n-m) \cdot c^2 \Bigg) ~ \Bigg]^{\frac{1}{2}}
\end{align}
where $x^i \in \hat{X}^{1,i}_{k|k-1}$, $x^j \in \hat{X}^{1,j}_{k|k-1}$ and $\Pi_n$ denotes the set of all permutations of size $m$ taken from the set $\{1,2,...,n\}$. The function $d_c(x,y) = \text{min}(c, \normvec{x-y}_2)$ where the parameter $c>0$ penalizes the cardinality mismatch between two sets. When $n<m$ Eq. (\ref{eq:ospa}) becomes $\Delta L^c_{k}(\hat{X}^{1,j}_{k|k-1},\hat{X}^{1,i}_{k|k-1})$. The above equation is called the optimal sub-pattern assignment (OSPA) \cite{Schuhmacher2008} of order 2. Then the \textit{cumulative tracking overlap score} for the time-window $[\kappa:K]$ is defined as:
\begin{align}
& Q_{\kappa:K}(s^i_{\kappa-1},s^j_{\kappa-1})= \\ 
&
\sum_{k=\kappa}^{K} \mathcal{I}(\mathcal{S}_a(s^i_{k-1}),\mathcal{S}_a(s^j_{k-1})) \cdot \Delta L^c_{k}(\hat{X}^{1,i}_{k|k-1},\hat{X}^{1,j}_{k|k-1}) \notag
\end{align}

\noindent where the function $\mathcal{I}(A,B)$ checks if the intersection of two regions $A$ and $B$ is non-empty and returns $1$, otherwise returns $\infty$. As we can see, the cumulative tracking overlap score will generate a low score if two agents track the exact same targets over a certain period of time. In other words when two agents have overlapping sensing ranges and they track the same number of targets with small positioning errors the cumulative tracking overlap score is minimized. Finally, in order to determine if there is tracking overlap between two agents over a time-window the cumulative tracking overlap score is tested against a pre-determined threshold $Q^{Th}$. If $Q_{\kappa:K} \le Q^{Th}$ then the two agents track with high certainty the same targets and thus one of them is removed from tracking. The removed agent generates a search plan and begins in the next time-step to search the surveillance region. 

\textbf{b) Control:} Finally the objective of tracking control is to find the optimal control action $u_k \in \mathbb{U}_k$ that must be taken at time step $k$ by each agent in order to maintain tracking of the detected targets. We should point out that the control action $u_k$ affect the received measurements $Z_k$ which in turn affects the multi-target state estimate $\hat{X}^1_k$ during the update step. Thus, ideally to optimize the control actions in this case it would require the knowledge of the future measurements. As a consequence, the objective function to optimize depends on future unknown measurements. Let this objective function be denoted as $\xi_\text{track}(u_k,Z_k)$.

Since, the future measurement set $Z_k$ is not available until the control action $u_k$ is applied, we generate the predicted measurement set $Z_{k|k-1}$ and we use it in place of $Z_k$. The predicted  measurement set $Z_{k|k-1}$ for each control action is generated as follows:
\begin{align}\label{eq:pims}
Z_{k|k-1} &=~ Z_{k|k-1} ~ \cup ~\{{\arg\max}_z ~  g_k(z|x,u_k)\} \notag \\
& \forall x \in \hat{X}^1_{k|k-1},~ \forall u_k \in \mathbb{U}_k
\end{align}
where $\hat{X}^1_{k|k-1}$ is the predicted multi-target state for the true targets. Thus the problem becomes:
$ u_{k}^{\star} = \arg\max~  \xi_\text{track}(u_{k},Z_{k|k-1})$. To optimize the track objective, the following steps are performed: The predicted SAT-PHD $D_{k|k-1}(x|Z_{1:k-1})$ is first computed from Eqn. (\ref{eq:satphd_predict}) without performing any control action. From this, we compute the predicted multi-target state $\hat{X}^1_{k|k-1}$ and for each admissible control action $u_k \in \mathbb{U}_k$ we generate the predicted measurement set $Z_{k|k-1}$ using Eqn. (\ref{eq:pims}). For each pair $(u_k, Z_{k|k-1})$ we perform a pseudo-correction step using Eqn. (\ref{eq:satphd_correct}) to produce the (pseudo) posterior SAT-PHD density $\hat{D}_k$. 

That said, we consider the information gain between the predicted $f_{k|k-1}(X|Z_{1:k-1})$ and the (pseudo) updated $\hat{f}_k({X|Z_{1:k-1},Z_{k|k-1},u_k})$ multi-target distributions as a measure of decreasing the uncertainty of the estimated multi-target state. The objective is then to maximize the information gain between the two multi-target distributions. To measure the information gain, we use as $\xi_\text{track}(.)$ the Renyi divergence \cite{Hero2008,Ristic2011,Ristic2010} which in our case is given by:
\begin{align} \label{eq:track_control}
&\xi_\text{track}(u_k,Z_{k|k-1}) =   \int_{\mathcal{X}^1} D_{k|k-1}(x) dx ~+~  \notag \\
& \frac{\alpha}{(1-\alpha)} \int_{\mathcal{X}^1} \hat{D}_{k}(x|Z_{k|k-1},u_k) dx ~-   \\
& \frac{1}{(1-\alpha)} \int_{\mathcal{X}^1} \hat{D}_{k}(x|Z_{k|k-1},u_k)^\alpha\notag D_{k|k-1}(x)^{1-\alpha} dx
\end{align}
where $0 < \alpha < 1$ determines the emphasis given on the tails of the two distributions. 


\begin{figure*}
	\centering
	\includegraphics[width=\textwidth]{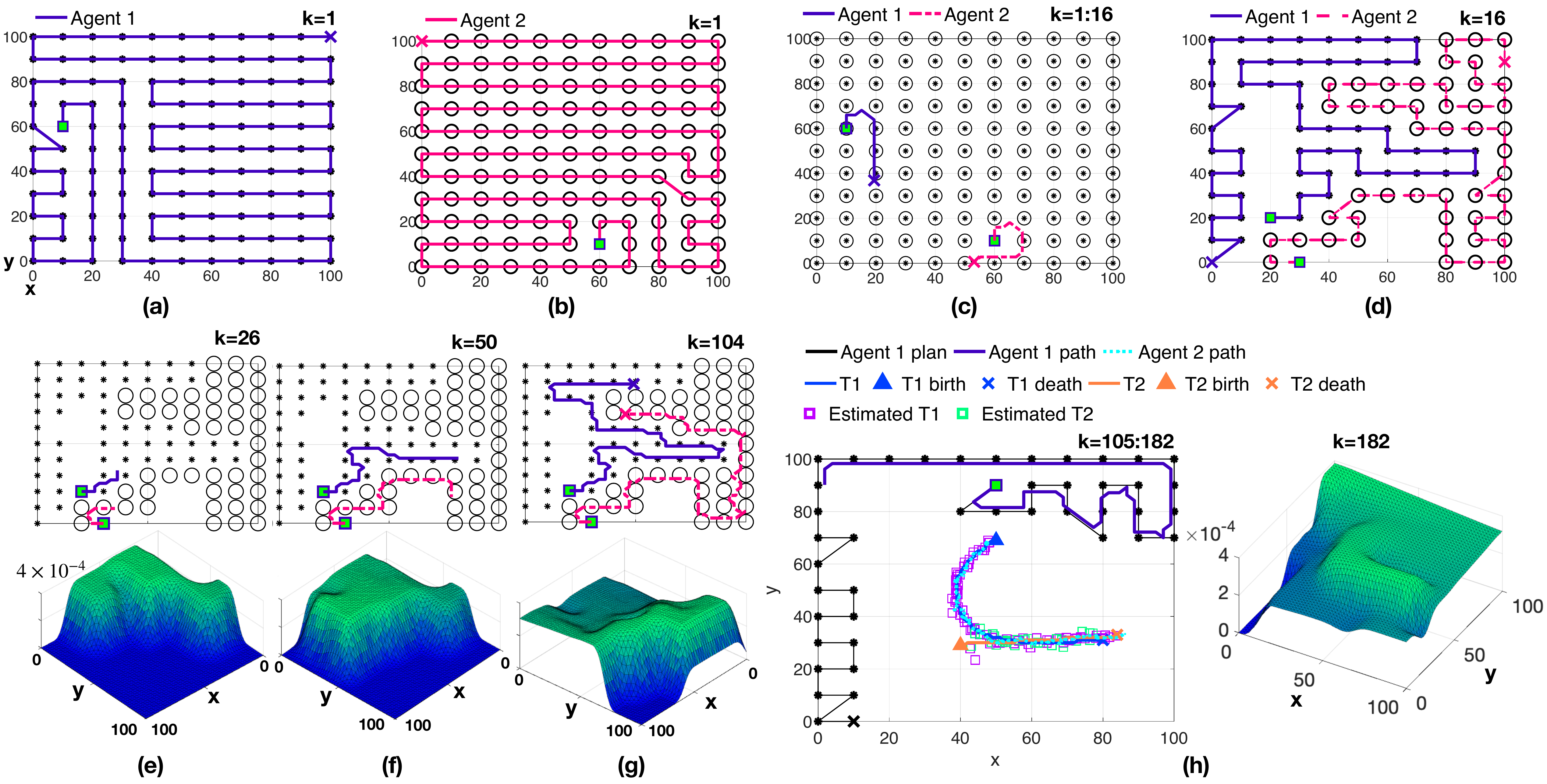}
	\caption{The figure shows the maneuvers of 2 agents for the task of search-and-track in a representative simulated scenario.}
	\label{fig:fig4}
\end{figure*}

%

\section{Evaluation}
\label{sec:Evaluation}
\subsection{Experimental Setup}

In our experimental setup we assume that the targets maneuver in an area of 100m $\times$ 100m. The target dynamics are modeled according to the near constant velocity model with the process noise being Gaussian. The single target transitional density is given by $\pi(x_k|x_{k-1}) = \mathcal{N}(x_k;Fx_{k-1},Q)$ where:
\begin{equation*}
	F = \begin{bmatrix} 
        1 & T & 0 & 0\\
        0 & 1 & 0 & 0\\
        0 & 0 & 1 & T\\
        0 & 0 & 0 & 1\\
       \end{bmatrix}, ~
    Q = \begin{bmatrix} 
        T/3 & T/2 & 0 & 0\\
        T/2 & T & 0 & 0\\
        0 & 0 & T/3 & T/2\\
        0 & 0 & T/2 & T\\
       \end{bmatrix}
\end{equation*}
with sampling interval $T=1$s. The target survival probability from time $k-1$ to time $k$ is constant $p_{s,k}(x_{k-1})=0.99$ and does not depend on the target's state. Once an agent detects a target it receives range and bearing measurements thus the measurement model is given by $h_k(x_k,s_k) = \left[\norm{s_k -\text{H}x_k}_2,~ \text{arctan}\left(\frac{s_y-\text{y}}{s_x - \text{x}}\right)\right]$ where $\text{H}$ is a matrix which extracts the target position from its state vector. The single target likelihood function is then given by $g(z_k|x_k,s_k) = \mathcal{N}(z_k;h_k(x_k,s_k),\Sigma^\top \Sigma)$ and sigma is defined as $\Sigma = \text{diag}(\sigma_\zeta,\sigma_\phi)$. The standard deviations $(\sigma_\zeta,\sigma_\phi)$ are range dependent and given by 	$\sigma_\zeta = \zeta_0 + \beta_\zeta \norm{s_k-\text{H}x_k}_2^2$ and $\sigma_\phi = \phi_0 + \beta_\phi \norm{s_k-\text{H}x_k}_2 $ respectively with $\zeta_0 = 1$m, $\beta_\zeta = 5\times10^{-3}\text{m}^{-1}$, $\phi_0 = \pi/180$rad and $\beta_\phi=10^{-5}\text{rad}/\text{m}$. 
Moreover, the agent receives spurious measurements (i.e. clutter) with fixed Poisson rate $\lambda_k = 10$ uniformly distributed over the measurement space. 
The agent's sensing model parameter $p_D^{\text{max}} = 0.99$ and the agent sensing area is $\mathcal{S}_{10}(s_k) = 10^2 ~\text{m}^2$.
The agent's dynamical model has radial displacement $\Delta_R=2$m, $N_R=2$ and $N_\theta = 8$ which gives a total of 17 control actions, including the initial position of the agent. The function $J_k(x)$ is constant and state independent and equal to $J_k(x) = 0.999~ \forall x, k$. The parameter $\alpha$ in Eqn. (\ref{eq:track_control}) is set to 0.5 and finally, the agent communication range is $C_R=50$m. In order to handle the non-linear measurement model, we have implemented a Sequential Monte Carlo version of the PHD filter \cite{Vo2003}. 

\subsection{Results}
A representative search-and-track scenario with 2 agents and 2 targets, which takes place during 200 time-steps is shown in Fig. \ref{fig:fig4}a - \ref{fig:fig4}h. In this scenario, 2 agents enter at $k=1$ the surveillance area of size $100\text{m} \times 100\text{m}$ at the locations marked with $\square$ in Fig. \ref{fig:fig4}a-\ref{fig:fig4}b with coordinates $(10, 60)$ and $(60, 10)$ for agents 1 and 2, respectively. The target birth/death times are $k=104/179$ and $k=144/182$ for targets 1 and 2, respectively. The target birth locations (marked with $\triangle$) are $(50, 69)$ and $(41, 29)$ for targets 1 and 2 respectively, and their corresponding death locations (marked with $\times$) are $(80, 31)$ and $(84, 33)$ as shown in Fig. \ref{fig:fig4}h. At each time-step, the agents in communication range cooperate in order to jointly search the surveillance area and track the detected targets. Otherwise, the agents optimize their individual objective and operate on their own. This is shown in Fig. \ref{fig:fig4}a - \ref{fig:fig4}b where the two agents are not in communication range and no targets are being estimated to exist inside their sensing range. 

More specifically, at $k=1$ agents 1 and 2 generate a search plan that they will use in order to traverse (i.e. search) the surveillance area. Note here that the produced search plans shown in Fig. \ref{fig:fig4}a - \ref{fig:fig4}b if executed, will visit every node (marked with $\star$ and $\circ$ for agents 1 and 2 respectively) in the search map and as a consequence the agents will search the whole surveillance region. Figure \ref{fig:fig4}c shows the execution of the aforementioned search plans during time-steps $k=1:16$ and the trajectories of agents 1 and 2 according to their dynamical models.

At time-step $k=16$ the two agents appear to be in communication range where they exchange and fuse their search densities and generate a joint search plan as discussed in subsection \ref{ssec:cooperative_search}. As a result the surveillance area that has not been searched so far is partitioned into two non-overlapping regions as shown in Fig. \ref{fig:fig4}d. In essence the joint search plan assigns the nodes $v \in \mathcal{V}$ which are associated with regions $r \subset \mathcal{A}$ to the two agents in such a way so that the overall area is searched as efficiently as possible. Thus during time-steps $k=17:104$, the two agents start executing the joint search plan as shown in Fig. \ref{fig:fig4}e - \ref{fig:fig4}g for time-steps $k=26$, $k=50$ and $k=104$. Fig. \ref{fig:fig4}e - \ref{fig:fig4}g also illustrates the fused search densities for the same time-steps. 
Next, at time-step $k=104$ target 1 is born inside agent's 2 sensing range. The agent estimates the presence of this target at $k=105$. As a consequence, agent 2 exits the joint search plan and begins to track target 1. Because at $k=104$ the two agents happen to be in communication range this information is transferred to agent 1 which recalculates its search plan to account for area dropped by agent 2. This is shown in Fig. \ref{fig:fig4}h. As you can observe, agent 1 recalculates its search plan which now includes nodes that have been initially assigned to agent 2. In addition, the same figure shows the search trajectory of agent 1 during time-steps $k=105:182$ and the tracking trajectory of agent 2 during the same period. At $k=144$ target 2 is born which is also being tracked by agent 2 during time-steps $k=147:182$. Between time-steps $k=147:179$, agent 2 tracks both targets as shown by the agent's trajectory and estimated target positions. Finally, the combined search density of the two agents for $k=182$ is shown (in this case, we have manually combined the two search densities in order to show the overall searched area, since the two agents are not in communication range at $k=182$).

\section{Conclusion} \label{sec:Conclusion}
In this work a novel decentralized cooperative multi-agent search-and-track framework has been proposed based on the theory of random finite sets (RFS). The Probability Hypothesis Density (PHD) filter has been extended to recursively propagate in time the search-and-track density which is used to produce cooperative searching and tracking strategies. The proposed framework is flexible and accounts for many of the challenges present in search and rescue missions including the unknown and time varying number of targets, the noisy sensor measurements, the uncertain target dynamics and the limited sensing range of the agents. Future work will focus on the real-world implementation and evaluation of the proposed framework.
\section*{Acknowledgments}
This work is supported by the European Union Civil Protection under grant agreement  No 783299  (SWIFTERS), by the European Union’s Horizon 2020 research and innovation programme under grant agreement No 739551 (KIOS CoE) and from the Republic of Cyprus through the Directorate General for European Programmes, Coordination and Development.

\flushbottom
\balance

\bibliographystyle{IEEEtran}
\bibliography{IEEEabrv,main} 

\end{document}